\documentclass[5p,12pt]{elsarticle}

\usepackage{amssymb}

\journal{Astroparticle Physics}

\begin{document}

\begin{frontmatter}



\title{Search for astrophysical high energy neutrino point sources with a False Discovery Rate controlling procedure}


\author{Bruny Baret$^1$, Mathieu Labare$^2$ and Daniel Bertrand$^2$}

\address{$^1$\emph{AstroParticule \& Cosmologie, CNRS/CEA/Universit\'e Paris Diderot/Observatoire de Paris, France}\\
  $^2$\emph{Interuniversity Institute for High Energy, Universit\'e Libre de Bruxelles, Belgique}}

\begin{abstract}
  A systematic multiple hypothesis testing approach is applied to the
  search for astrophysical sources of high energy neutrinos. 
  The method is based on the  maximisation of the detection power maintaining the control of the confidence level of an hypothetical discovery. This is
  achieved by using the so-called "False Discovery Rate" (FDR)
  controlling procedure. It has the advantage to be independent of the
  signal modelling and to naturally take into account the trial factor. 
  Moreover it is well suited to the detection of multiple sources.
\end{abstract}


\end{frontmatter}


\section{Introduction} 
With the construction of the IceCube detector\cite{icecube} in the ice of the South Pole glacier following 10 years of data taking by its predecessor AMANDA\cite{amanda} and the completion of Antares \cite{antares} in the depth of the Mediterranean sea opening the way to its kilometric successor Km3NeT\cite{km3net}, high energy neutrino astronomy is entering a new era.\\
As neutrinos are neutral and only sensitive to weak interaction they can directly be related to their production source. Their detection would be a proof of the existence of hadronic processes associated to the most violent objects of the Universe and could provide the so far missing ``smoking gun'' for the acceleration of high energy cosmic ray protons.
The neutrino telescopes are based on the detection of Cherenkov light emitted by the secondary lepton produced by the charged current interaction of neutrinos in a large volume ($\sim 1~km^3$) of transparent medium (ice or sea water) using an array of photomultipliers. The present analysis focuses on the detection of muons produced by charged current interaction of high energy ($ > 100~GeV$) neutrinos.
The muon track direction is reconstructed by using the arrival time of Cherenkov photons and the muon energy is estimated using the number of triggered photomultipliers. 
The neutrino direction accuracy depends on the kinematics of the reaction and the intrinsic detector resolution  (of the order of $1^{\circ}$ in the ice for AMANDA and $0.1^{\circ}$ in the water for Antares).\\
These detectors have to cope with a physical background produced by the interactions of high energy cosmic rays with the atmosphere. The prompt
muons are suppressed by using the Earth as a shield while the atmospheric neutrinos constitute an irreducible background. The aim of the present analysis is to identify spatially localised excesses of events on top of this background.\\
In this paper, an original approach is presented to increase the sensitivity to high energy neutrino point source signal, while controlling the confidence level of an hypothetical discovery in a model independent way. We will focus on time integrated northern sky surveys, looking for sources independently of known electromagnetic counterparts. The current statistical approaches and their limitations as well as the basics of the False Discovery Rate controlling procedure are presented in section 2. The application of FDR to the search of point sources with the AMANDA neutrino telescope is presented in section 3 and the performances of the method in section 4.


\section{Statistical tools for point source search}

\subsection{The multiple hypothesis tests}
Looking for a point source in a sky of events can be seen as multiple hypothesis tests. Indeed, events which are described by some random variable $X$ can either come from background and follow a statistical law $H_{0}$ called the null hypothesis, or be signal events following an alternative $H_{1}$ law. A testing procedure consists in defining a so-called critical region which is a sub-region of the  possible values of $X$ where the null hypothesis will be rejected and hence the related events considered as sources. Outside this region, events will be considered as background. The possible different outcomes are summarized in Table \ref{outcomesumup}. \\

\begin{table}[h] 
 \begin{center}
\begin{tabular}{|c|c|c|c|}
\hline
~	& 	$H_0$	&	$H_1$	&Total\\
\hline
\hline
rejected	& 	$U$	&	$T$	&$R$  \\
\hline
not rejected	&	$V$	&	$S$	&$m-R$\\
\hline\hline
Total		&	$m_0$	&	$m-m_0$	&$m$  \\
\hline
\end{tabular}
\end{center}
\caption{Summary of the different outcome counts of multiple hypothesis tests.}
\label{outcomesumup}
\end{table}

Several ``quality'' indicators for a given testing procedure can be constructed from these numbers, first by the number of Type I errors $U$ which are the (false) rejections of true nulls. The expectation of $U$ is called the size of the test. The most widely used related quality indicator of a testing procedure is the Family Wise Error Rate ($FWER$) which is defined as the probability of making at least one Type I error:
\begin{equation}
FWER=Pr(U>0)
\end{equation}
This indicator might appear inadequate in case a source is made up of several rejected events where it will lead to overconservative tests.
Another indicator, that we will use in the following, is the False Discovery Rate ($FDR$) defined as the expectation value of the rate of false rejections among all rejections:\\
\begin{equation}
FDR=E(\frac{U}{R \vee 1})
\end{equation}
A simple interpretation is that the expectation value of the confidence level of a discovery will be $1-FDR$, since $FDR$ is an estimator of the size of the test.
Another important quantity is the number of Type II errors $S$, the non-rejection of the signal. The probability of not making such an error at each test defines the power of the procedure. One is generally interested in finding the testing procedure which maximizes the power for a given size. Technically, such a procedure will consist in finding a variable to test and defining the critical region where the null hypothesis will be rejected. 
In most cases the critical region will be characterized by setting an adequate threshold on this variable.
\subsection{Classical approaches and their limitations}
When both $H_0$ and $H_1$ are known, the most powerful procedure is the Neyman-Pearson test which relies on likelihood ratio \cite{neymanpearson}. One defines:
\begin{equation}
\lambda=\frac{\mathcal{L}(x_1,...,x_N|H_0)}{\mathcal{L}(x_1,...,x_N|H_1)}
\end{equation}  
where $\mathcal{L}$ is the likelihood function for the $N$ outcomes $x_i$ of the tested random variable $X$ considering hypotheses $H_{j}$ with $j\in \{0,1\}$.
The critical region will be defined by $\lambda\le c_\alpha$ where $c_\alpha$ is such that if one desires a confidence level of at least $1-\alpha$:
\begin{equation}
Pr(\lambda(x)\le c_\alpha | H_0 )= \alpha
\end{equation}
This is what is commonly used in neutrino astronomy \cite{madison}\cite{ific}. But there are important drawbacks.\\
The first problem of this approach is its dependence on signal models since one has to know the  $H_1$ probability density function (pdf)  making the method strongly model dependent. 
The second is that the determination of $c_\alpha$ is often not explicitly computable and needs to be evaluated using heavy Monte Carlo simulations. 
Indeed in practice when one wants to insure a high confidence level the critical region will be characterized by very low statistics. One usually overcomes this problem by extrapolating the distribution of higher p-values.  Moreover, instead of looking for a realization below an {\it a priori} fixed threshold, one usually looks for the highest excess in data and then checks the probability to obtain a lower p-value, adapting this way the number of Monte-Carlo equivalent experiments to the data. 
The problem here is that it is not suited to the detection of multiple sources. 
\\
Alternatively a model independent ``blind'' analysis \cite{blind} could be performed using only the background hypothesis which can be inferred from the data. 
This method avoids the use of  heavy Monte Carlo simulations for the determination of the critical region. 
Furthermore, fixing {\it a priori} the number of false nulls does not limit the multiple source detection capability.\\
In the following we will use a set of $N$ realizations $\{x_{i}\}_{i\in \{1,..,N\}}$ of a random variable $X$, and its pdf  $P_{0}(x)$ under the null hypothesis $H_{0}$. 
The first step is to compute p-values $p_{i}$  for each realization $x_{i}$, defined as the probability to observe at least $x_{i}$ under the null hypothesis:\\
\begin{equation}
p_{i}=\int_{x_{i}}^{\infty} P_{0}(x)\;\mathrm{d}x
\end{equation}
A naive way to perform the test would be to reject all realizations with p-value lower than a fixed threshold $p_t=\alpha$. But after $N$ tests the final confidence level of any discovery will be lower than $(1-p_t)\times N$ in the most general case, and $1-(1-p_t)^N$ in the particular case of independent tests. This is the trial factor effect which we would like to take into account without large numbers of heavy Monte-Carlo simulations.\\
A very conservative way to overcome this  and hence to control the FWER is the so-called Bonferonni approach \cite{bonferonni}. It consists in dividing the individual threshold by the number of trials (the equivalent for the independent case is the Sidak \cite{bonferonni} procedure which replaces the threshold by $(1-(1-p_t)^\frac{1}{N}$). But then the detection power decreases very strongly since the individual minimum threshold p-value will be divided by $N$, leading to an increase of the number of type II errors.\\
We will describe in the following a procedure which controls the FDR and hence the confidence level, and which by relaxing the FWER control makes it possible to reach a high detection power.

\subsection{False Discovery Rate controlling procedure}
This FDR procedure has been first developed in \cite{Benjamini1} and applied to several kinds of astrophysical and cosmological searches \cite{miller}\cite{millerb}\cite{pires}. It is based on an adaptive threshold on p-values which guarantees both high detection power and control of the FDR and hence the confidence level of an hypothetical detection.
The procedure itself is rather simple. Given a set of $N$ p-values and a FDR input value $\alpha$: 
\begin{itemize}
\item sort the p-values in increasing order, to get the ordered set $\{p_{i}\}_{i\in {1,..,N}}$.
\item find $i_{c}$ so that 
\begin{equation}
i_{c}=\max( i\in \{1,..,N\} ~|~ p_{i}\le\frac{1}{\chi}.\alpha.\frac{i}{N} )
\end{equation}
where $\chi$ is a coefficient that has to be introduced to account for dependency between the tests. It is generally 1 and the theoretical value in the worst case is $\sum_{i=1}^{N}\frac{1}{i}$. The latter value makes the procedure more conservative but is rarely necessary. In the particular case of p-values distributed uniformly the coefficient is unity \cite{benjamini2}.
\item all tests with index $i\le i_{c}$ will reject the Null Hypothesis and we considered as sources.  
\end{itemize}
The procedure is proved to be "minimax" (optimal choice minimizing risk in decision theory) for Gaussian, Poissonian and exponential distributions \cite{donoho}. Moreover, unlike the classical approach it is well suited for the detection of multiple sources since it relies on the rate of false discoveries.\\



\section{Application to AMANDA reconstructed high energy neutrino skies}
\label{sec:amandatest}
The FDR method described in what follows is based on the use of a random variable built as a convolution of 2 variables: the angular position and the energy of the neutrino candidates. It has been tested on event distributions corresponding to the high energy neutrino northern sky as seen by AMANDA \cite{amandadata}. These data have been used to determine the atmospheric neutrino background sample needed to construct the Null Hypothesis pdf. On the other hand a cosmic neutrino sample was used to derive limits and discovery potential corresponding to one year of data taking as well as to compare with the classical method based on Likelihood ratio.
\subsection{Null Hypothesis}
\label{sec:nullhyp}
\subsubsection{Information from angular position}\label{sec:fdrpos}
\begin{figure}[!h]
\includegraphics[width=90mm]{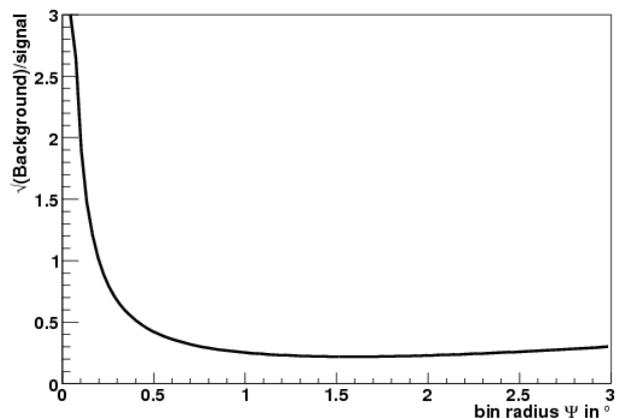}
\caption{\small Evolution of $r_{\sqrt{b}/s}$ versus the radius of the bin for a resolution of $1^{\circ}$.}
\label{fig:ratiosigbgd}
\end{figure}

Due to the detector cylindrical symmetry along the north-south axis, the distribution of the angular position is uniform in right ascension and therefore only depends on the declination $\delta$. 
Given the low statistics accumulated annually by the AMANDA detector (of the order of 1000 events), the number of neutrinos in a sky area of the order of the detector resolution $\rho$ ( $\sim 3\deg$ ) follows a Poisson statistic. 
 For each reconstructed neutrino in this area corresponds a circular search region for which the space angle radius $R$ will vary as a function of $\delta$ in order to ensure the same Null Hypothesis all over the sky.\\
The detector resolution is defined as the median of the point spread function (PSF) which is the distribution of the differences between the true neutrino direction and the reconstructed one.
Neglecting small dependency on the neutrino energy, the angular resolution is assumed to be the same for signal and background events.
The mean number of expected background events in a region of angular radius $R_{opt}(\delta)$ is defined as:
\begin{equation}
\mu_{bg}(\delta)=\frac{1-\cos R_{opt}(\delta)}{2\cos \delta \sin R_{opt}(\delta)}\int_{\delta-R_{opt}(\delta)}^{\delta+R_{opt}(\delta)} f(\theta)\;\mathrm{d}\theta 
\end{equation}
where $R_{opt}(\delta)$ is the angular radius of the region minimizing the square root of background over signal ratio $r_{\sqrt{b}/s}$, corresponding \cite{sigbak} to $1.585\times\rho$ .
If $\mu_{max}$ is the maximum value of $\mu_{bg}(\delta)$ over the sky, in order to ensure the same background event count $\mu_{max}$ all over the sky $R$ must be:
\begin{equation}
R(\delta)=Arccos(1-\frac{\mu_{max}}{\mu_{bg}(\delta)} (1-\cos R_{opt}(\delta) ))
\end{equation}
This way $R$ will be at least as large as $R_{opt}$. This is the best compromise since for a circular area the ratio $r_{\sqrt{b}/s}$ increases slower for values higher than $R_{opt}$ than for lower ones as can be seen in Fig \ref{fig:ratiosigbgd} which represents the evolution of $r_{\sqrt{b}/s}$  with the  radius for an hypothetical PSF of $1^{\circ}$ resolution.


\subsubsection{Information on energy}
The energy is another observable discriminating signal and background, since the cosmic spectrum is expected to be generically harder than the atmospheric one. 
Cosmic neutrino energies are supposed to follow a power law of index $-2$ versus $-3.7$ for atmospheric secondaries. The most robust and simple variable available to evaluate the neutrino energy is the number of hits $N_{ch}$, defined as the number of optical modules triggered by the photons emitted by the muon when passing through the detector \cite{amandareco}. Like the background angular position distribution, the pdf $P_0(N_{ch})$ corresponding to this variable can be extracted from the data (Figure \ref{fig:PDF-Nch}).  It is combined with the angular position distribution to lead to the total energy accumulated in a search region of the sky.\\
In the case of a single neutrino the only information is its own energy which pdf is $P_0$. In the case where there is only one neighbour neutrino, the total energy distribution will be :
\begin{equation} 
P_1(x) = \sum_{a=0}^x P_0(a)\,P_0(x-a) 
\end{equation}
By recursion, the total accumulated energy within a search region centered on a neutrino and containing $n$ extra neutrinos is distributed as :
\begin{equation} 
P_n(x) = \sum_{a=0}^x P_0(a)\,P_{n-1}(x-a) \end{equation}
These probabilities have to be weighted by the Poissonian distribution $\mathcal{P}(n;\mu_{max})$ to observe $n$ neutrinos in a region where $\mu_{max}$ are expected. The total energy distribution corresponding to the Null Hypothesis is: 
\begin{equation}
P_{TOT}(N_{ch}) = \sum_{n=0}^{\infty}\mathcal{P}(n;\mu_{max})\,P_n(N_{ch})
\label{eq:total_nch}
\end{equation}

\subsection{Clustering}
\label{clustering}
The method can lead to multiple counting of signal events after the application of the FDR procedure as the region attached to each neutrino can overlap. 
One way to overcome this is to perform a spatial clustering on rejected
neutrinos. The most suitable scheme here is a hierarchical minimal
distance clustering \cite{clustering}. The algorithm can be summarized
as follows:\\
Initially each neutrino of the rejected set ${N_r}$ is attached to a different cluster $C_i$.\\
The following procedure is then applied as long as the number of clusters is decreasing.\\
If the distance between two clusters is lower than the sum of the resolutions $R(a)$ and $R(b)$ on the direction of the corresponding neutrinos, the two clusters they belong to are merged\\
$$\mbox{if}\ \ d_{ij} < R(a) + R(b)\\
\mbox{then}\ \ C_i = C_j = C_i \cup C_j$$
where $$d_{ij} =\{ \min (distance(\nu_a , \nu_b)) \ | \  \nu_a \in C_i \, , \nu_b \in C_j \} $$

At the end of the clustering, each cluster will be considered as a signal source. The false discovery rate estimator does not account for individual neutrinos anymore but for clusters.

\subsection{Final cut}\label{finalcut}
The procedure at this stage does not yet allow perfect control of the false discovery rate at high source luminosity. This can be understood from the principle of the FDR rejection, which is designed to detect tiny excesses with respect to the Null Hypothesis. Hence a source with high potential of detection obviously deviates from the main purpose of the method. This generates artifacts appearing as small clusters mainly made up of a single neutrino. For consistency, we want the method to control the False Discovery Rate even in these extreme cases. The simplest and most efficient way is to cut off single neutrino clusters when at least another bigger cluster exists within the same events sample. This does not affect the efficiency of the method at low luminosity.


\section{Results }
	
The method has been tested on simulation samples where neutrino point sources have been added to an atmospheric background of 1000 events corresponding to the statistics collected by AMANDA in one year \cite{amanda20002003}.
The statistical estimators of false discovery rate and detection power are defined as the mean value of each quantity over a large number of artificial sky realizations. Given the binomial nature of these estimators (a discovery is either true or false) the statistical error $\epsilon$ will be \cite{metzger}:
\begin{equation} \epsilon = \sqrt{\frac{x(x-1)}{N-1}} \end{equation}where $x$ is the estimator and N the number of sky realizations used to determine the mean.
With  $10^3$ sky realizations one reaches a statistical error of order $0.22\%$ which is sufficient to test a $99.5\%$ confidence level control. The following section first describes the variables and pdfs used to define the Null Hypothesis and to build background and source Monte Carlo generator. The false discovery rate control by the procedure is then checked and discovery potential of the method is quantified.

\subsection{Generation of the background test sample}
\begin{figure}[!h]
\includegraphics[width=90mm]{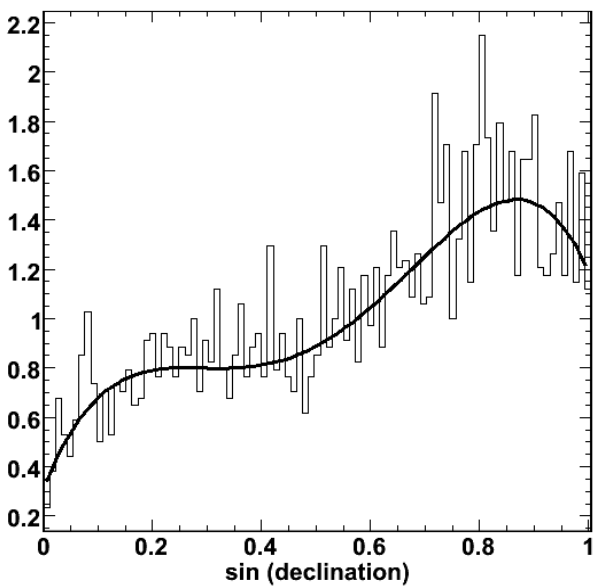} 
\caption{\small PDF of the sine of the declination of events observed with the AMANDA detector. }
\label{fig:PDF-Background}
\end{figure}

As described in section \ref{sec:nullhyp}, the two observables used by the method are the angular position of events and their energy.
Hence the distributions of the declination and of the energy of the real events were fitted with analytical curves. 
The expected background as a function of the sine of the declination is shown in Figure \ref{fig:PDF-Background}. It has been fitted with a polynomial function.

\begin{figure}[!h]
\includegraphics[width=90mm]{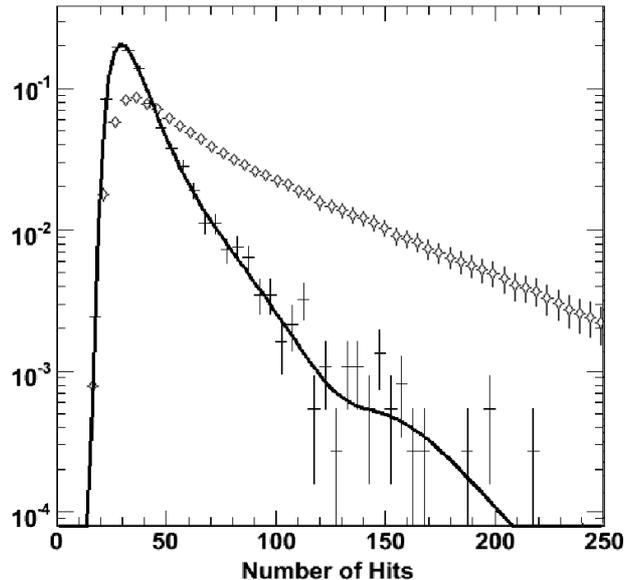} 
\caption{\small Normalized distributions of the number of hits for the AMANDA detector. Points are data (and hence considered as purely atmospheric) and the line is the fit of the corresponding probability density function. Empty squares correspond to a simulation of the detector response to a cosmic flux.}
\label{fig:PDF-Nch}
\end{figure}

The number of hits ($N_{ch}$) triggered by the passing muon is  used as energy estimator. Its distribution is shown in Figure \ref{fig:PDF-Nch} together with the fit used for the Monte-Carlo generation.

 

Since the detector is symmetric around the north-south axis, the right ascension distribution is uniform.
These three pdfs are used to generate the Monte Carlo realizations of background skies. 

\subsection{Generation of the test source sample}
Neutrino sources were generated on the basis of the detector response corresponding to an energy distribution following a power law of index $-2$. These events were submitted to the full reconstruction chain of AMANDA. The energy distribution of these events is represented in Fig. \ref{fig:PDF-Nch}.
For all tests, one source is simulated with luminosity varying between 0 (pure background) and 20 neutrinos, at various declinations to be compared with an average expected background count of two. In order to take into account the detector resolution, cosmic neutrinos are spread around the source location following the PSF at this declination.


\begin{figure}[!h]
\includegraphics[width=90mm]{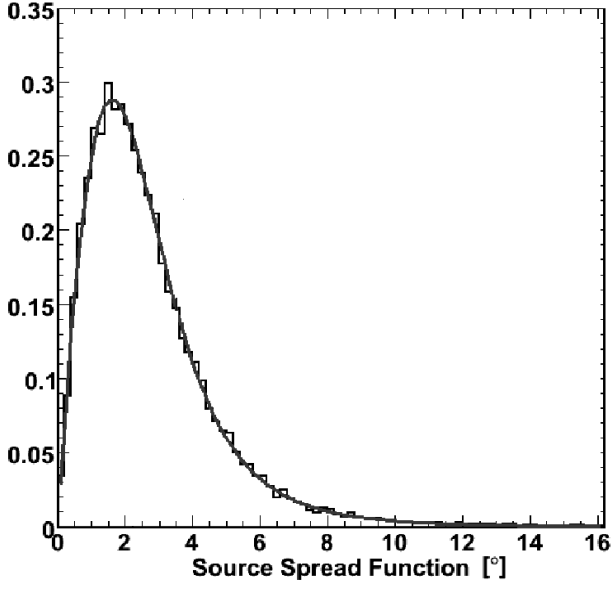} 
\caption{\small Point spread function at 22.5$^{\circ}$ of declination. The curve shows the fit function used for MC generation. }
\label{fig:SSF_22_5}
\end{figure}

\begin{figure}[!h]
\includegraphics[width=90mm]{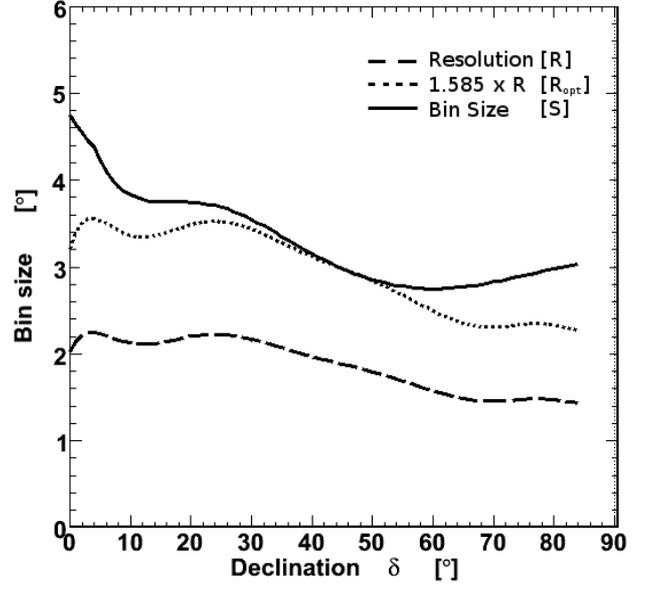}
\caption{\small Detector resolution $R$ , optimal signal to noise ratio bin size $R_{opt} $ and the optimal constant background bin size $S$ which is eventually used as bin definition. }
\label{fig:Binsize}
\end{figure}

An example of point spread function can been seen in Figure \ref{fig:SSF_22_5} and the evolution of its characteristic width in Figure \ref{fig:Binsize} . 
\begin{figure}[!h]
\includegraphics[width=90mm]{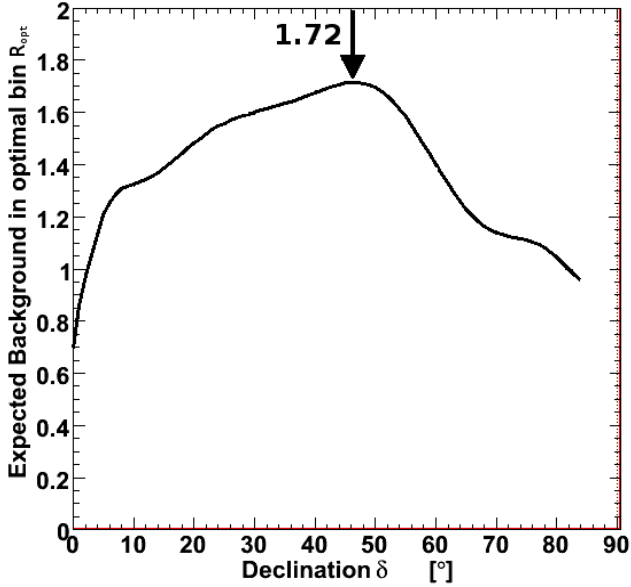}
\caption{\small Expected background in optimal bin size $R_{opt}$. Maximal value is used as expected background to build bin size in the whole sky.  }
\label{fig:MaxExpBg}
\end{figure}

\subsection{Construction of the Null Hypothesis pdf}
\begin{figure}[!h]
\includegraphics[width=90mm]{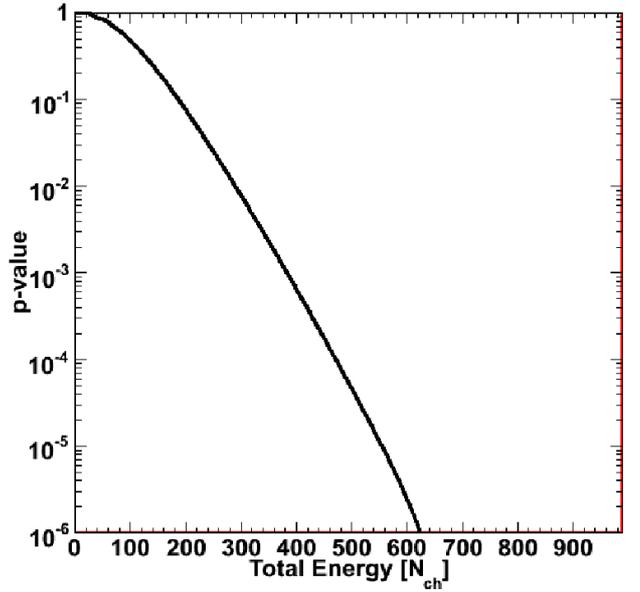}
\caption{\small Cumulative Distribution Function on Total number of hits expected for a pure background sky with 1000 atmospheric neutrinos  }
\label{fig:CDF-ttl}
\end{figure}

As mentioned in section \ref{sec:fdrpos}, a circular search region of angular radius $R_{opt}$ is centered on each reconstructed neutrino. 
The expected number of events (Fig.\ref{fig:MaxExpBg}) is computed thanks to the known background density.
The search region radius  $R(\delta)$ is then defined in such a way that it contains always the maximal expected background of 1.72 neutrino independent of its angular position. The variation of this radius as a function of the declination can be seen in Figure \ref{fig:Binsize}.
The pdf of the Null Hypothesis is then computed from  Eq.\ref{eq:total_nch}. The corresponding cumulative density function is shown in Figure \ref{fig:CDF-ttl}.

%

\subsection{FDR control}
\begin{figure}[!h]
\includegraphics[width=90mm]{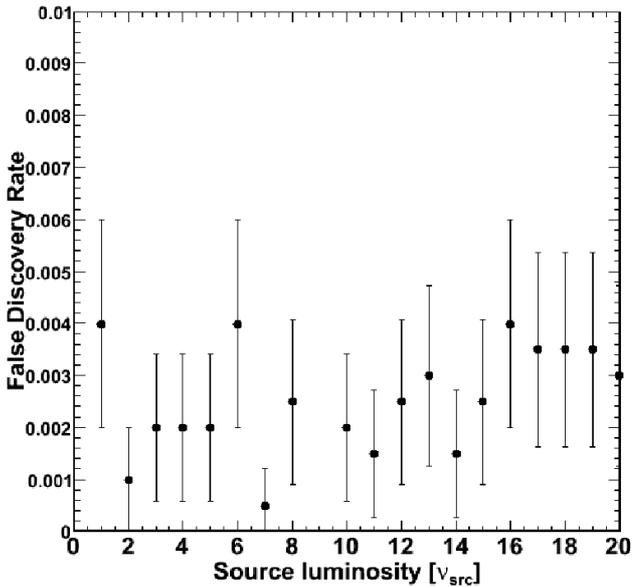}
\caption{\small False Discovery Rate obtained with 1000 atmospheric neutrinos versus luminosity of a source located at $22.5^{\circ}$ declination and requested FDR of 0.5\%.}
\label{fig:FDR_Lum}
\end{figure}

The false discovery rate control was checked at ten different source declinations varying between $0^{\circ}$ and $85^{\circ}$ and with source luminosities ranging from $0$ to $20$ neutrinos. The false discovery rate is well controlled in all configurations at the requested $99.5\%$ confidence level. An example can be seen in Figure \ref{fig:FDR_Lum} for a source located at $22.5^{\circ}$ of declination. 

\begin{figure}[!h]
\includegraphics[width=90mm]{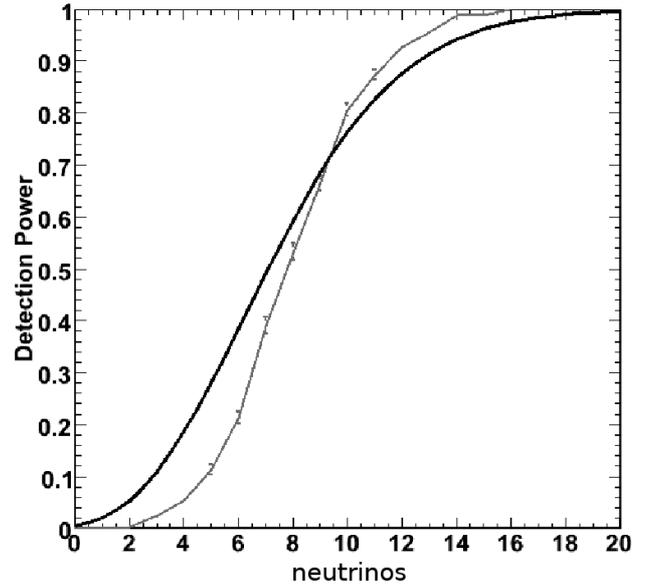}
\caption{\small Detection probability of a source located at $22.5^{\circ}$ above a background of 1000 atmospheric neutrinos versus source luminosity. Grey curve corresponds to discrete number of detected neutrinos while the black one corresponds to a Poisson averaged mean luminosity. }
\label{fig:DP_Lum}
\end{figure} 

\subsection{Discovery potential} 
The probability to identify the source is estimated on the same test sample. Results are shown in Figure~\ref{fig:DP_Lum} for a source located at $22.5^{\circ}$ of declination. The grey curve corresponds to the detection probability  $\mathcal{DP}_{ev}(\nu_{src})$ of a given number of detected neutrinos from the source $\nu_{src}$. The number of detected events emitted by a source of a given luminosity follows a Poisson distribution with a mean depending on the source flux. Consequently the detection power (black curve in Fig.\ref{fig:DP_Lum}) as a function of the luminosity $\bar\nu_{src}$ is computed by summing the probabilities of detecting $n$ neutrinos weighted by the Poisson probability $\mathcal{P}(n|\bar\nu_{src})$.
\begin{equation}
\mathcal{DP}_{src}(\bar{\nu}_{src}) = \sum_{n=0}^{+\infty} \mathcal{P}(n|\bar{\nu}_{src}) \times \mathcal{DP}_{ev}(n)  
\end{equation}

\begin{figure}[!h]
\includegraphics[width=90mm]{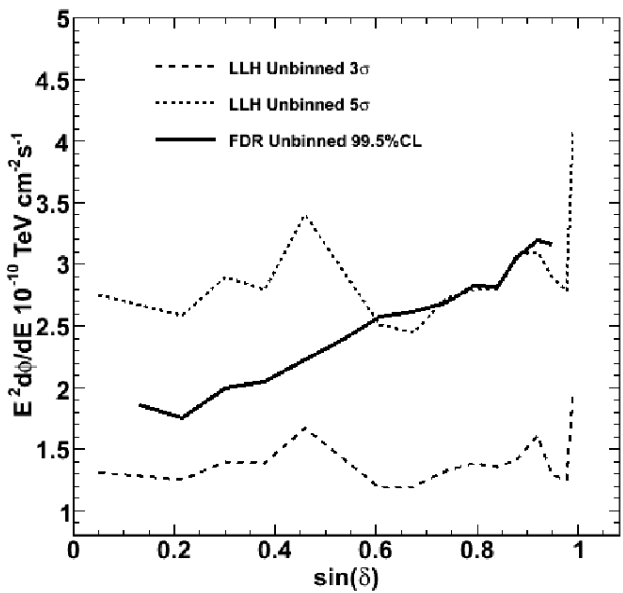}
\caption{\small Flux discovery potential for a detection efficiency of 50\% after 1 year of data taking }
\label{fig:DP50}
\end{figure}

\begin{figure}[!h]
\includegraphics[width=90mm]{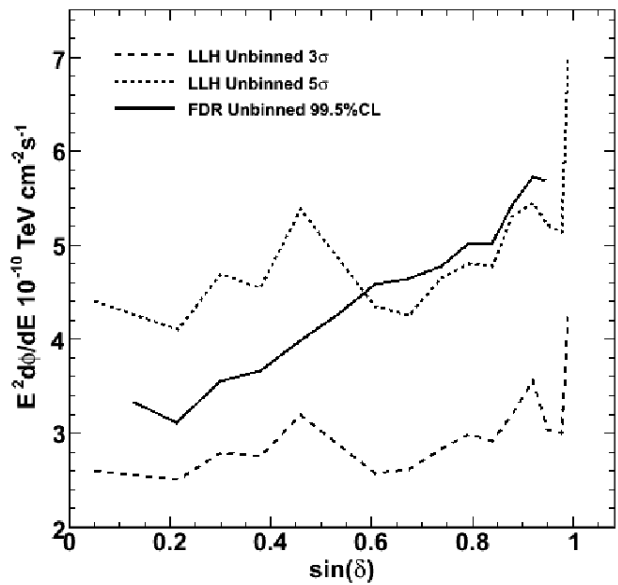}
\caption{\small Flux discovery potential for a detection efficiency of 90\% after 1 year of data taking   }
\label{fig:DP90}
 \end{figure}

To get physically meaningful information these numbers of source neutrinos have to be translated into fluxes by performing a convolution with the effective area of the detector. 
The fluxes needed to have 50$\%$ (Fig.\ref{fig:DP50}) and 90$\%$ (Fig.\ref{fig:DP90}) chance to detect a source with a confidence level of $99.5\%$ for one year of data-taking were computed. For instance, as can be seen in Fig.\ref{fig:DP_Lum}, the minimal mean source strength is 7.1 cosmic neutrinos over 1.72 background neutrinos to have 50$\%$ probability to identify the source at a declination of $22.5^{\circ}$. 
This number increases to 12.6 cosmic neutrinos for a detection probability of 90$\%$.  

The comparison of these results with other methods is not straightforward since, the FDR is able to handle multiple sources while other methods, \emph{eg.} likelihood ratio maximisation are optimized for a flux model and only deal with the highest excess on a sky. However, it is reasonable to compare log likelihood ratio (LLH) method corresponding to a $5\sigma$ excess above background and FDR procedure at $99.5\%$CL\@. Indeed the trial factor effect for a statistic of $1000$ neutrinos in a sky will give a probability of approximately $0.5\%$ that an excess corresponding to at least $5\sigma$ would appear on a pure background sky \cite{jbraunthesis}.


For comparison, fluxes corresponding to detection efficiencies of $50\%$ and $90\%$ are presented in Fig.\ref{fig:DP50} and \ref{fig:DP90}, together with the corresponding curves for LLH for  $3\sigma$ and $5\sigma$ excesses.
It has to be underlined that due to the trial factor effect an excess of $3\sigma$ or higher would appear with a probability of $64\%$ in a pure background sky of 1000 events.
It can be seen that the FDR method is clearly more sensitive than the LLH for a  $5\sigma$ effect at low declinations. This behaviour is linked to the increase of background in this region of the sky which implies for the LLH method an increase of the source luminosity to maintain the same detection efficiency.

\subsection{Limits and sensitivity}
\begin{figure}[!h]
\includegraphics[width=90mm]{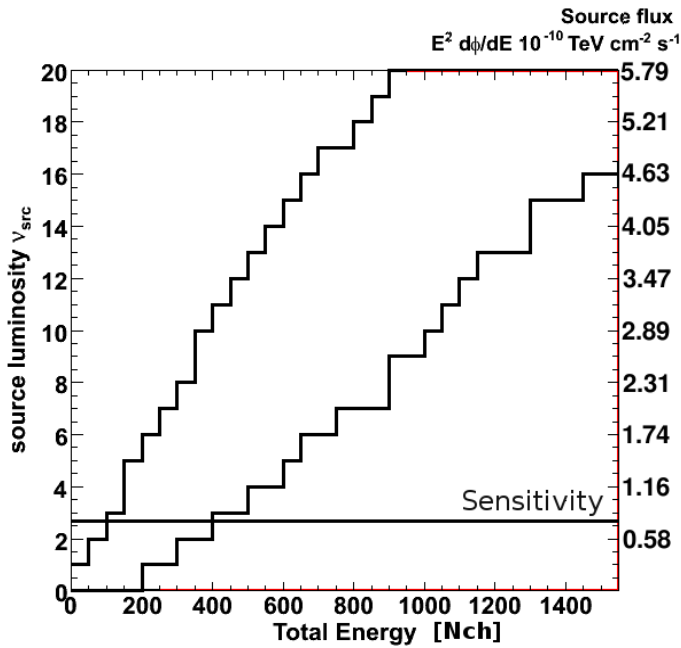}
\caption{\small Feldman-Cousins confidence belt for declination 22.5$^{\circ}$ after 1 year of data taking }
\label{fig:FC}
\end{figure}

\begin{figure}[!h]
\includegraphics[width=90mm]{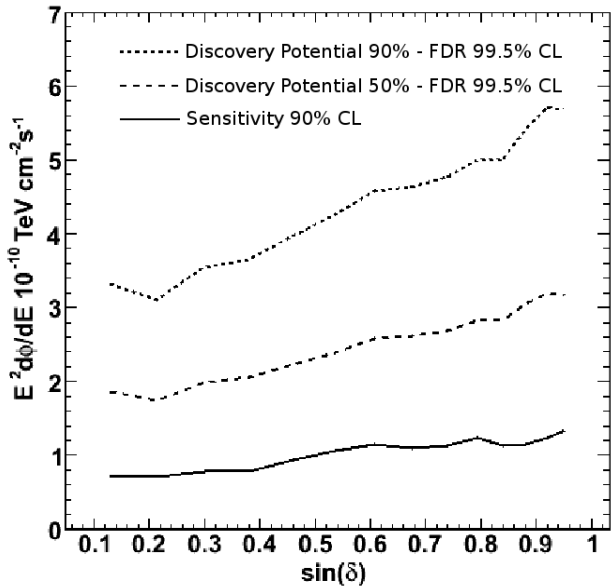} 
\caption{\small Sensitivity and discovery potential after 1 year of data taking }
\label{fig:Sensitivity}
\end{figure}

We use the unified ordering algorithm proposed by Feldman and Cousins \cite{feldmancousins} to construct confidence belts and quantify the sensitivity of the FDR procedure. Sensitivity is defined here as the average upper limit obtained from confidence belts for all possible experimental outcomes weighted by their probability if no signal is present.\\
An example of a confidence belt is shown for a possible source located at $22.5^{\circ}$ declination in Figure \ref{fig:FC}. The sensitivity shown in this figure corresponds to a differential cosmic neutrino flux of $0.77\times10^{-10}$ TeV cm$^{-2}$ s$^{-1}$.  The variation of the sensitivity as a function of the declination is presented in Figure \ref{fig:Sensitivity}.


\section{Summary and conclusions}
We have presented an original statistical approach to the search of
point-like sources of cosmic neutrinos with Cherenkov telescopes, which
allows to control the confidence level of an hypothetical discovery with multiple hypothesis test. 
Its main advantages are that it is independent of any model of the signal production, that it takes into account the trial factor effect and that, unlike existing methods, it naturally includes the search for multiple sources.
Its performances were compared with a likelihood ratio minimisation approach using the data collected by the AMANDA neutrino telescope. In the case of the detection of a single source in the context of a particular flux model, the FDR method results in a sensitivity equivalent to the Bayesian approach, showing even slightly better performances at low declination or in case of a very faint source.


\end{document}